\documentclass{aims2}
\usepackage{amsmath}
  \usepackage{paralist}
 \usepackage[colorlinks=true]{hyperref}
\hypersetup{urlcolor=blue, citecolor=red}

\newcommand{\dd}{\delta}
\newcommand{\bn}{{\bar n}}
\newcommand{\bm}{{\bar m}}
\newcommand{\bbm}{{\bar M}}
\newcommand{\bw}{{\bar w}}

\newcommand{\hgg}{{\hat\gamma}}

\newcommand{\be}{\begin{equation}}
\newcommand{\ee}{\end{equation}}
\newcommand{\ph}{\varphi}

\newcommand{\e}{\varepsilon}
\newcommand{\Om}{\Omega}
\newcommand{\po}{{\partial\Omega}}
\newcommand{\pd}{{\partial D}}
\newcommand{\ZZ}{\mathbb{Z}}
\newcommand{\RR}{\mathbb{R}}
\newcommand{\RRR}{\RR^3}

\newcommand{\bc}{\setminus}
\newcommand{\GG}{\Gamma}
\newcommand{\Lo}{{L\bc\{0\}}}

\newcommand{\p} {\partial}

\def\currentvolume{X}
 \def\currentissue{X}
  \def\currentyear{200X}
   \def\currentmonth{XX}
    \def\ppages{X--XX}
     \def\DOI{10.3934/xx.xx.xx.xx}
     
\newtheorem{theorem}{Theorem}[section]

\newtheorem{lemma}[theorem]{Lemma}
\newtheorem{proposition}[theorem]{Proposition}

\theoremstyle{definition}

\newtheorem{remark}{Remark}

\title[ Surface Energy]
      {Continuum Surface Energy from a Lattice Model}

\author[Phoebus Rosakis]{}

\subjclass{Primary: 74Qxx; Secondary:74N05,  11P21.}
 \keywords{Atomistic models, continuum models, pair potential,  surface energy,  lattice point problem}

 \email{rosakis@tem.uoc.gr}

\begin{document}
\maketitle

\centerline{\scshape Phoebus Rosakis }
\medskip
{\footnotesize
 \centerline{Department of  Applied Mathematics}
   \centerline{University of Crete
}
   \centerline{Heraklion 70013, Greece}
} 

\bigskip


\begin{abstract}
 We investigate  connections between the continuum and atomistic descriptions of deformable crystals, using certain interesting results from number theory. The energy of a deformed crystal is calculated in the context of a lattice model with general binary interactions in two dimensions. A new bond counting approach is used, which reduces the problem to the lattice point problem of number theory. The main contribution is an explicit formula for the  surface energy density  as a function of the deformation gradient and boundary normal. The result is valid for a large class of domains, including faceted (polygonal) shapes and regions with piecewise smooth boundaries.
\end{abstract}

\section{Introduction}

This article is concerned with the derivation of continuum surface energy from a standard lattice model, by exploiting results related to certain lattice point problems of number theory, e.g. \cite{barany,CCD,huxley,ivic,Pick}. 

We study the energy of a crystal, modelled as the part of a Bravais lattice $L$ contained in a reference region $\Om\subset\RR^d$, with atoms (elements of $\Om\cap L$) interacting through a pair potential $\ph$. The potential may have unrestricted range but must decay fast enough. The crystal is subjected to a smooth deformation $y\colon\Om\to\RR^d$. The energy under consideration is 
\be\label{-4}E\{\Om,y\}=\sum_{x\in\Om\cap L}\;\sum_{z\in(\Om\cap L)\bc x}\ph\left(|y(z)-y(x)|\right)\ee
To approach  the continuum limit, one may scale the lattice, i.e., replace $L$ by $\e L$ and rescale the potential to $\ph_\e=\ph(\frac{\cdot}{\e})$,  then study asymptotics of the energy  as $\e\to0$ \cite{Blanc,Mora}. Equivalently,  one can  rescale the region to $r\Om$ and the deformation to $y_r=ry(\frac{\cdot}{r})$, with $r=1/\e$, but leave $L$ and $\ph$ unscaled.  

 The aim of this paper is to write the discrete energy (1.1) in the canonical form of continuum mechanics,  with emphasis on its dependence on the geometry of the boundary $\po$.
We describe our main results, Propositions \ref{conpo} and \ref{cure}. For the case $d=2$, suppose $\Om$ is a convex region whose boundary is piecewise smooth and  may contain crystallographic facets (subject to certain restrictions)  with outward unit normal $n$, and that the deformation is homogeneous, $y(x)=Fx$, $x\in\Om$ for some\footnote{$M^{2\times2}_+$ is the set of $2\times2$ matrices with positive determinant.}  $F\in M^{2\times2}_+$.  Proposition \ref{conpo} shows that the energy \eqref{-4} satisfies
\be\label{-3}E\{k\Om,y\}=\int_{k\Om }W(F) dx +\int_{k\po} { \hgg}(F,n)ds+o(k),\ee
as $k\to\infty$, $k\in\ZZ$. Here $W(F)$ is given by the Cauchy-Born formula (\eqref{0} below) and $k\Om=\{z\colon z=kx,  x\in\Om\}$ is the dilated region. The  new aspect of this result is the explicit computation of the \emph{surface energy density function} $\hgg$; see \eqref{three} below. It turns out that the dependence of $\hgg(F,n)$ on the normal $n$ involves a dense set of discontinuities (Proposition \ref{propg}). As a result, the hypotheses of the  standard surface energy minimization theorem yielding the Wulff shape,  may not be fulfilled in general  \cite{dac},   \cite{fonseca}. This pathological behavior   is due to geometrical reasons, stemming from the difference between the continuum volume $|\Om|$ and  what is sometimes termed the ``discrete volume'' $\#(\Om\cap L)$, cf. \cite{CCD} (associating a cell of unit volume to each atom in the discrete body $\Om\cap L)$.  These difficulties are resolved in Proposition \ref{cure}, where  the energy is written in the alternative form
\be\label{-2}E\{k\Om,y\}=\int_{\Om(k) }W(F) dx +\int_{\po(k)} { \gamma_\circ}(F,n)ds+o(k),\ee
as $k\to\infty$, $k\in\ZZ$. 
 Here $\Om(k)$ is a suitably defined region containing \emph{the same lattice points} as $k\Om$, namely $\Om(k)\cap L=k\Om\cap L$, thus  the discrete energy is the same. On the other hand, $\Om(k)$ is constructed so that its continuum and discrete volumes coincide to order $o(k)$. Specifically, when $\Om$ is a lattice polygon (i.e., its vertices are lattice points), then $\Om(k)$ is the rational polygon obtained by translating each side of $k\Om$ outwards by half the distance to the next crystallographic plane with the same normal.  However, $\Om(k)$ is not a dilation of  $\Om$ in general. In this case, the dominant part of the $o(k)$ term in  \eqref{-2} is an $O(1)$ corner energy that is obtained exactly. Our  main contribution is the following \emph{explicit formula for the surface energy density} $\gamma_{\circ}\colon M^{2\times2}_+\times S^1\to\RR$:
\be\label{-1} \gamma_\circ(F,n)=-\frac{1}{4}\sum_{w\in \Lo} |w\cdot n|\ph(|Fw|).\ee
Unlike $\hgg$ in  \eqref{-3}, $\gamma_\circ(F,\cdot )$ is Lipschitz on $S^1$, thus it abides by the hypotheses of the Wulff theorem.
 
On the other hand, if $\Om$ is a smooth $C^2$ strictly convex region,  on may choose $\Om(k)=k\Om$; moreover \eqref{-2} then holds for any real (not only integer) sequence $k\to\infty$; see Proposition \ref{cure}.
 
Formula \eqref{-1} for the surface energy density is analogous to the well-known Cauchy-Born formula for the stored energy function in the first term of \eqref{-2}:
\be\label{0}W(F)=\frac{1}{2}\sum_{w\in \Lo} \ph(|Fw|).\ee

The first rigorous derivation of continuum energy functions from atomistic models is due to Blanc, Le Bris and Lions \cite{Blanc}, who study (among other problems)  the asymptotics of  the energy\footnote{The energy in \cite{Blanc} is divided  by $\#(\Om\cap\e L)$ and has  rescaled potential $\ph_\e$.} of a crystal $\Om\cap\e L$, subject to a prescribed smooth deformation $y:\Om\to\RRR$ as $\e\to 0$. The dominant term is the usual elastic energy $\int_\Om W(\nabla y(x))dx$ with $W$ given by \eqref{0}. The next term, of order $\e$ in Theorem 3 of \cite{Blanc}, is a surface integral over $\po$; the integrand depends on  the deformation gradient and the geometry of $\po$. The form of this surface energy is not explicit. Terms of order $\e^2$ include a volume integral of an explicitly determined higher gradient energy, but also surface terms; the latter are left unspecified. 

As shown in one dimension by Mora-Corral \cite{Mora}, the higher order terms in the asymptotic expansion of the energy in powers of $\e$ depend on the choice of the sequence of $\e\to 0$.  In Theorem 3 of \cite{Blanc}, this choice is restricted by the hypothesis that there exist a sequence $\e=\e_k\to 0$ as $k\to\infty$, such that  $\#(\Om\cap\e_kL)=|\Om|/\e_k^d$ (in dimension $d$). Letting $r=1/\e$ and scaling $\Om$ instead of the lattice, this  means that for some sequence $r_k\to\infty$, 
\be\label{bbl}\#(r_k\Om\cap L)=|r_k\Om|.\ee
In the present work we rely on  bond counting arguments instead of asymptotics to a large extent. A byproduct of this approach is an explanation of this sequential dependence issue. Our method hinges on finding, for each $w\in L$, the \emph{$w$-bond number of $\Om$}:
\be
\label{nwom}N_w(\Om)=\#\{x\colon x\in\Om\cap L, \; x+w \in\Om\cap L\},\ee
that is, the number of pairs of atoms within $\Om$ separated by a given vector $w$.
For large $r$, the dominant contribution to $N_w(r\Om)$ is $\#(r\Om\cap L)$.  Finding the asymptotics of $\#(r\Om\cap L)$ as $r\to\infty$  is 
 the \emph{lattice point problem} of number theory \cite{CCD,ivic,Tsang}.  This reduces to studying the  \emph{lattice point remainder} $R(r)=\#(r\Om\cap L)-|r\Om|$, the difference of the two sides of \eqref{bbl}. In the  context of crystals, letting each atom  have unit mass, and since the lattice cell in $\ZZ^d$ has unit measure, the continuum mass density should equal $1$. Thus $R(r)=\#(r\Om\cap L)-|r\Om|$ is the difference of the discrete  and the continuum mass of the body, which do not coincide in general. This causes problems with the continuum notion of mass density; see  Remark \ref{rem7.5}.  Another difficulty is that  fixing the (discrete) mass of the body in surface energy minimization problems is not equivalent to the traditional volume constraint.
 
In two dimensions, the problem of characterizing $R(r)$ is open for general domains with piecewise $C^1$ boundary, while even the \emph{Gauss circle problem}  ($\Om$ the unit disk, $L=\ZZ^2$)  is not completely settled  \cite{huxley}. Through \eqref{nwom}, the lattice point remainder enters  our estimates for the energy $E\{r\Om,y_r\}$, whose asymptotic form thus depends on the sequence $r_k$ through $R(r_k)$. This can be problematic as   $R$ is discontinuous and  highly oscillatory.  In general, the behavior of $R (  r)$ 
depends strongly on the shape of $\po$. For $\Om$ a lattice polygon (one whose vertices are lattice points), $R(r)$ is of same order as the surface energy---$R(r)=O(r^{d-1} )$ in dimension $d$---and can be characterized explicitly; see, e.g., Lemma \ref{pickvar} below. For smooth convex domains in $\RR^2$,  as shown by van der Corput \cite{C}, $R( r )=O(r^{2/3})$, between the orders of the surface and the gradient energy of \cite{Blanc}, but  difficult to characterize   \cite{huxley}. 

Hypothesis \eqref{bbl} made by \cite{Blanc} is equivalent to existence of a sequence $r_k$ such that $R(r_k)=0$, thus it eliminates such undesirable higher order terms from a Riemann sum of the elastic energy. In addition, it has the desirable property that the continuum and discrete mass coincide for $r_k\Om$.  Unfortunately however, it is not known for which choices of  $\Om$ such a sequence exists.   Another approach  (see Theil \cite{Theil}) is to define the surface energy  as the difference of  the discrete energy $E\{r\Om,y\}$ and a bulk energy of the form $\frac{\#(r\Om\cap L)}{|r\Om|}\int_\Om W(F)$. Under the hypotheses of  Proposition \ref{conpo}, for $r\in\ZZ$ we find  that the resulting surface energy  equals the one in \eqref{-2}  and is thus free from the pathology associated with $\hat\gamma$ in \eqref{-3}.  On the other hand,   the lattice point number $\#(r\Om\cap L)$ cannot be obtained explicitly for general  piecewise $C^1$ domains, as discussed above.  This means that  the bulk energy is not explicitly characterized in this approach. Here we choose to maintain the standard notion of continuum bulk energy, $\int_\Om W(F)$, without the additional discrete factor. This issue is discussed further in Remark \ref{rem10}.  
 
We summarize our results. Crystals typically occur in faceted form in their natural state (for instance, the Wulff shape, e.g., \cite{herring,dac,fonseca}). This is because of surface energetics affecting crystal growth, but also because cleavage fracture creates new surfaces along special crystallographic planes. This means that they can be modeled as \emph{crystallographic polyhedra}, whose facets inhabit crystallographic planes (that contain a two-dimensional sublattice of $L$). 
 
 In Section \ref{sec2} we assume that $\Om$ is a \emph{lattice polytope}, i.e, one whose vertices are lattice points. This does not sacrifice too much generality over crystallographic polytopes. Indeed, if $\Om$ is crystallographic polytope, then $k\Om$ is a lattice polytope for some $k\in\ZZ$. In addition, there is a lattice polytope $\Om'\subset\Om$, such that $\Om'\cap L=\Om\cap L$. If $\Om$ is convex, then $\Om'=\hbox{conv}\{\Om\cap L\}$. In view of Theorem 3 of \cite{Blanc}, one expects that the dominant surface energy term does not involve higher gradients of the deformation. Accordingly, it suffices to assume that the deformation is homogeneous (affine). To keep the geometry simple, we confine our analysis to two dimensions. 
 Unlike \cite{Blanc,Mora}, initially we do not employ a limit process, but rather a \emph{bond counting} technique. The computation of the energy is reduced to that  of \eqref{nwom}. We then show that this calculation reduces to a number of lattice point problems. The solution of the latter for lattice polygons is furnished by Pick's Theorem  \cite{Pick}.  The lattice point remainder $R(k)$ is known exactly (for $k\in\ZZ$) and contributes to the surface energy explicitly, being of the same order.
 
 In Section \ref{sec3} we compute the energy of polygonal crystals. 
 For an interatomic potential of finite but arbitrary range, we obtain the energy of essentially any convex lattice polygon \emph{exactly} (Proposition \ref{finener}). This result is not asymptotic and does not suffer from the sequential dependence issue explored in \cite{Mora}. Let the deformation be $y(x)=Fx$, $x\in\Om$. The energy equals the \emph{exact} sum of the elastic energy $\int_\Om W(F)dx$ plus the \emph{surface energy} $\int_\po \gamma_\diamond(F,\bn)dx$, plus the \emph{corner energy} $\sum_{i=1}^N \tau(F,n_i,n_{i-1})$, summed over the $N$ vertices of $\Om$. The surface energy density is explicitly obtained:
\be\label{one}
\gamma_\diamond(F,\bn)=-\frac{1}{4}\sum_{w\in \Lo}\frac{1}{|\bn|} \left( |w\cdot \bn|-1\right)\ph(|Fw|),
\ee
where $\bn$ is a normal to $\po$ whose components on each facet are the Miller indices (irreducible integers) of the corresponding lattice plane, and $\ph$ is the interatomic potential. 
The corner energy $\tau(F,n_i,n_{i-1}) $ is also explicit but more complicated; apart from $F$, it depends on the two unit normals of the facets meeting at the $i$th vertex.

For an infinite range potential this result retains only asymptotic validity for a lattice polygon $k\Om$ as  $k\to \infty$; the three energies just mentioned are the first three terms of the asymptotic expansion of the energy for large $k$ (Proposition \ref{infener}). 

In Section \ref{sec4}, we consider regions with smooth boundaries. Because of its construction based on lattice polygons, the surface energy density \eqref{one} is only defined for ``rational'' directions of the surface normal; $n=(\nu_1,\nu_2)\in S^1$ is called \emph{rational} if $\nu_2/\nu_1$ is a rational number or $\nu_1=0$, \emph{irrational} otherwise. It is natural to ask how \eqref{one} can be extended to irrational normals. When $\Om$ is strictly convex and $\po$ is smooth for example, the normal is irrational almost everywhere on $\po$. We start by letting $\po$ be of class $C^2$ with positive curvature. The key observation is that the \emph{convex hull} of all lattice points contained in such an $\Om$ is a lattice polygon. This allows us to use number-theoretic results on the asymptotic properties of such hulls due to B{\'a}r{\'a}ny and Larman \cite{barany}; see also the survey \cite{ivic}. Perhaps surprisingly, the surface energy density for smooth strictly convex regions (Proposition \ref{smopo}) is \emph{different} from \eqref{one}. It is given by \eqref{-1},
where $n$ is  the \emph{unit normal} to $\po$ and can take on irrational values. The difference is due to the lattice point remainder  $R(  r)=O(r^{2/3})$ \cite{C,huxley}, which is now of lower order than the surface energy.  As a result, the  asymptotic expression for the energy of inflated regions $r\Om$ is \emph{sequence-independent};  the sequence of $r\to\infty$ is not restricted to be integer but arbitrary. 

We then consider more general  regions  with piecewise $C^1$ boundary that comprises flat facets as well as curves with positive curvature. For such regions, the surface energy density function, now defined for all $n\in S^1$, is obtained in Proposition \ref{conpo}:
\be\label{three}\hgg(F,n)
=\begin{cases} 
\gamma_\diamond(F,\bn),
& n \;\;\hbox{rational} \;\;(\bn/|\bn|=n),
\\ 
\gamma_\circ(F,n),&n\;\;\hbox{irrational},
\end{cases}
\ee
 with $\gamma_\diamond$ from \eqref{one} and $\gamma_\circ$ from \eqref{-1}. The dependence of the surface energy density on the  normal is rather pathological. Specifically, $\hgg(F,\cdot)\colon S^1\to\RR$ is continuous at irrational $n$, discontinuous at rational $n$, and almost nowhere differentiable (Proposition \ref{propg}). Because of this, the surface energy density need not satisfy the usual hypotheses of the Wulff theorem (determining the domain that minimizes the surface energy under fixed measure); see e.g. \cite{dac,fonseca}, but also  Remark \ref{rem6}. 
 
 In Section \ref{sec5}, we resolve the difficulties due to  discontinuous dependence of the surface energy on the unit normal. This dependence is due to the behavior of the lattice point remainder of regions with rational boundary normal.   
 We  alter the region $\Om$ so as to change its measure, but not the lattice points it contains. The goal is that the lattice point remainder of the modified region should be of lower order than the surface energy.   For example,  if $\Om$, hence $k\Om$, is  a lattice polygon, translate each side of $k\Om$ outwards by half the distance to the next crystallographic plane with the same normal.  This results in a \emph{rational polygon}  $\Om(k)$ that contains the same lattice points as $k\Om$. Note, however, that $\Om(k)$ is not a rescaling of $\Om$ in general.  The lattice point remainder  of  $\Om(k)$ is $O(1)$ as $k\to\infty$,  of lower order than the surface energy. This allows us  to write the latter in the form 
$ \int_{\po(k)}\gamma_\circ(F,n) ds$.
The associated surface energy density $\gamma_\circ$, given by \eqref{-1}, is  Lipschitz continuous in the unit normal. This and additional considerations discussed in Section \ref{sec5}, show that \emph{$\gamma_\circ$ is the appropriate  density for the determination of the Wulff  shape} that minimizes the  surface energy  $ \int_{\po}\gamma_\circ(F,n) ds$ over a suitable class of regions $\Om$ with fixed measure \cite{herring,dac,fonseca}.  

A more realistic approach to surface energy would allow for  ``relaxation'' of atomic positions from the macroscopic deformation near the boundary. Such deviations might  be determined by minimization of the atomistic energy. This is a formidable problem in the present setting (more than one dimension, general boundary geometry, arbitrary interaction range, nonconvex potentials).   One of the few results in this direction is due to Braides and Cicalese \cite{bracic}; they obtain the relaxed surface energy in one dimension using $\GG$-convergence. The result is not explicit and seems difficult to compare quantitatively with the explicit ``constrained" energy of Mora-Corral \cite{Mora}.   In  two dimensions, Theil (\cite{Theil}, Theorem 1.4) calculates the relaxed surface energy of a crystal with quadratic short range potentials; 
 the result is in the form of a perturbation of the constrained surface energy.
 
In order to obtain  quantitative information on the difference between the relaxed and constrained surface energies, numerical optimization of the atomistic energy was recently performed for a completely unconstrained, Lennard-Jones two-dimensional crystal \cite{rosak}. Atomic positions were allowed to relax from initial positions forming a lattice triangle or hexagon with low Miller-index boundary. The constrained energy was obtained by minimizing over the deformation gradient matrix of a homogeneous deformation that the atoms are constrained to follow. It was found that the difference between the relaxed and constrained surface energies is typically less than three percent (after the appropriate scaling and bulk energy is accounted for). This suggests that  in some situations  the relaxed and constrained surface energies may be quite close. In analogous one-dimensional computations, the results agree qualitatively with the conclusions of \cite{bracic}, while the difference between the relaxed and constrained surface energies is less than one percent.   Values of this difference computed in three dimensions using density-functional theory  for low Miller-index surfaces in various metals are usually less than three percent; see, e.g.,  \cite{ehf}.
 
Many of the results presented here, in particular expressions \eqref{one} through \eqref{three} for the surface energy density, are valid for three-dimensional crystals as well \cite{rosak}. 
 \section{The Bond Counting Approach}\label{sec2}

For subsets $P$, $Q$ of $\RR^n$, define the Minkowski sum $P\oplus Q=\{p+q:p\in P,\; q\in Q\}$ and write $p+Q=\{p\}\oplus Q$. The \emph{lattice} is $L=\ZZ^2$ unless otherwise noted. 

\begin{remark}\label{rem100} All of our results can be immediately adapted to any Bravais Lattice $L^*$ by incorporating the linear mapping from $L$ onto $L^*$ into the deformation. Expressions like \eqref{-1} remain valid if $L$ is replaced by $L^*$, provided the  linear mapping from $L$ onto $L^*$ has unit Jacobian determinant. \end{remark}
For $x=(\alpha,\beta)\in \ZZ^2$ let 
 $$\gcd(x)=\gcd(|\alpha|,|\beta|), \quad \bar x=\frac{1}{\gcd(x)}x,\quad x^\perp=(\beta,-\alpha).$$
 We assume that the reference region $\Om\subset \RR^2$ is a \emph{convex body}, or a compact convex set with nonempty interior.
Fix $w\in L$, let $x\in L$ and define $b=b(x,w)=\{z\in\RR^2\colon z=x+tw,\; 0\le t\le 1\}=\hbox{conv}\{x,x+w\}$ as the \emph{bond starting at $x$ with bond vector $w$} . The set 
\be\label{bw}B_w(\Om)=\{b:b=b(x,w),x\in \Om\cap L,x+w\in\Om\cap L\}\ee 
is the set of all $w$-bonds of $\Om$ (bonds with bond vector $w$). We will use the abbreviation
$$b_0=b(0,w).$$
 The energy of the homogeneous deformation $y(x)=Fx$ can be written as
\be\label{eeenergyy} E\{\Om,y\}=\frac{1}{2}\sum_{x\in\Om\cap L} \sum_{\substack{w\in \Lo\\x+w\in \Om\cap L}}\ph(|Fw|).
\ee
The factor of $1/2$ occurs since $b(x,w)=b(x+w,-w)$ and the potential $\ph$ is even in $w$. Interchanging the order of summation above we obtain
\begin{equation*}\begin{split}E\{\Om,y\}&=\frac{1}{2}\sum_{w\in \Lo} \sum_{\substack{x\in \Om\\x+w\in \Om\cap L}}\ph(|Fw|)\\
&=\frac{1}{2}\sum_{w\in \Lo} \,\sum_{b\in B_w( \Om)}\ph(|Fw|)
=
\frac{1}{2}\sum_{w\in \Lo}\ph(|Fw|) \sum_{b\in B_w( \Om)}1.\end{split}
 \end{equation*}
Evidently, in order to determine the energy, it suffices to calculate, for each $w\in L$, the \emph{$w$-bond number of $\Om$}, i.e., $N_w( \Om)=\#B_w(\Om)$; see \eqref{bw}:
\be\label{eey} E\{\Om,y\}=
\frac{1}{2}\sum_{w\in \Lo}\ph(|Fw|)N_w( \Om).
\ee
Clearly the number of $w$-bonds ``starting'' in $\Om$ equals the number of lattice points of $\Om$:
$$\#\{b=b(x,w): x\in\Om\cap L \}=\#(\Om\cap L) .$$
Some of these bonds are not contained in $B_w(\Om)$:
\be\label{t11}
N_w(\Om)=\#(\Om\cap L) -\# T_w(\Om),\quad T_w(\Om)= \{b=b(x,w)\colon x\in \Om\cap L,x+w\not\in \Om\cap L\}.
\ee
For $w\in L$ let 
\be\label{spp}
 S^+_w=\{x\colon x\in\po,\; x+w\not\in\Om\}.
\ee
so that $S_w^+$ is the part of $\po$ through which $w$ points outwards.
Denote by $T^\dagger_w(\Om)$ the set of all $w$-bonds that intersect $S^+_w$ and terminate outside $\Om$.
\be\label{t2}T^\dagger_w(\Om) =\{b:b=b(x,w)\in B_w(L), b\cap S^+_w\neq\emptyset, x+w\not\in\Om\}.\ee
 Some of these bonds ``straddle'' $\Om$, that is, have both endpoints outside $\Om$ but intersect $\po$; specifically,
\be\label{t3}T^\ddagger_w(\Om)=\{b(x,w)\in T^\dagger_w(\Om): x\not\in\Om,x+w\not\in\Om\}.
\ee
 Then obviously in view of \eqref{t11},
$$T_w(\Om)=T^\dagger_w(\Om)\bc T^\ddagger_w(\Om).$$
As a result,
\be\label{t4}N_w(\Om)=\#(\Om\cap L) -\# T^\dagger_w(\Om)+\# T^\ddagger_w(\Om).\ee
Roughly speaking, the number of $w$-bonds in $\Om$ equals the number of lattice points in it, minus the number of bonds that traverse the boundary at least once, plus the number of bonds that traverse the boundary twice. The reason for the splitting \eqref{t4} is that each term can be evaluated using results from geometric number theory.

 One important case we will consider is when $\Om\subset \RR^2$ is a \emph{convex lattice polygon}. In particular, $ \Om=\hbox{conv}\{v_1,\ldots,v_N\}$, the convex hull of its $N$ \emph{vertices} $v_i\in L$, $i\in\{1,\dots,N\}$, which are lattice points. The boundary $\po$ consists of $N$ \emph{facets} $S_i=\hbox{conv}\{v_i,v_{i+1}\}$, where $v_{N+1}=v_1$ and $S_{N+1}=S_1$. Letting $m_i=v_{i+1}-v_i$, $\bar m_i=m_i/\gcd(m_i)$, the \emph{Miller normal} $\bar n_i$ to $S_i$ is $\bar n_i=\bar m_i^\perp$, so that $\gcd({\bar n_i})=1$. 
The number of lattice points in $\Om$, $ \#(\Om\cap L)$, is addressed by Pick's Theorem, \cite{Pick,Reeve,CCD}, a variant of which is the following

 \begin{lemma}\label{pickvar} Let $\Om$ be a simple closed lattice polygon with facets $S_i$ and outward Miller normal $\bn=\bn_i$ on $S_i$. Then
 \be\label{pick1} \#(\Om\cap L)=|\Om|+\frac{1}{2}\sum_{i=1}^N \frac{ |S_i|}{|\bn_i|}+1 . \ee
 Equivalently, letting $\theta_i$ be the (dihedral) angle between normals of facets meeting at the $i$th vertex,
 \be\label{cpick} \#(\Om\cap L)=\int_\Om 1 dx+\int_\po\frac{1}{2|\bn|} ds+\sum_{i=1}^N \frac{\theta_i}{2\pi}.\ee
\end{lemma}

\begin{proof} Pick's Theorem \cite{Pick,Reeve} states that 
\be\label{pickk}|\Om|= \#(\mathring\Om\cap L)+\frac{1}{2}\#(\po\cap L)-1= \#(\Om\cap L)-\frac{1}{2}\#(\po\cap L)-1\ee
(since $\Om$ is closed). If two neighboring lattice points in a facet $S_i$ differ by $\bar m_i\in L$ (with relatively prime components), then $\#(S_i\cap L)=|S_i|/|\bar m_i|+1$ while $\#(\po\cap L)=\sum_{i=1}^N [\#(S_i\cap L) -1]= \sum_{i=1}^N|S_i|/|\bar m_i|$ since each $S_i$ contains both its endpoints. Now the Miller normal $\bn_i=\bar m_i^\perp$, so that $ |\bn_i|=|\bar m_i|$ and \eqref{pick1} follows. Also, \eqref{cpick} is a trivial consequence of \eqref{pick1}, given that the sum in \eqref{cpick} equals $1$. 
\end{proof}

\begin{remark}\label{rem5}  The shape of naturally occurring crystals is very often faceted (polyhedral).  Thus one might start by assuming that $\Om$ is a polygon, though not necessarily a lattice polygon. In that case though, there is a lattice polygon $\Om'$ such that $\Om\cap L=\Om'\cap L$.  For example, if $\Om$ is convex, let $\Om'=\hbox{conv}\{\Om\cap L\}$. If $\Om$ is a crystallographic polygon, so that its facets are contained in crystallographic lines, then   its vertices need not be lattice points. However, one can then show that there is some integer $k$ such that $k\Om$ is a lattice polygon. 
\end{remark}

\begin{remark}\label{rem1}Eq. \eqref{cpick} has an interesting interpretation. It exactly equates a discrete quantity (number of atoms in $\Om$) with a continuum expression: the ``volume'' integral of a bulk density, plus the ``surface'' integral of a surface density, plus contributions of corners. We will show in the sequel that both the $w$-bond number $N_w(\Om)$ and the energy admit analogous representations.\end{remark}
 
Recall that $S=\partial\Om$ consists of $N$ facets $S_i$, $i=1,\ldots, N$, each with unit normal $n_i$, outward with respect to $\Om$.   Define
\be\label{sp}
J(w)=\left\{i\in \ZZ: 1\le i\le N, n_i\cdot w>0 \right\}, \quad S^+_w=\bigcup_{i\in J(w)} S_i,
\ee
The first term in \eqref{t4} is given by \eqref{pick1}. Turning to the second term, let $P_i(w)$ be the parallelogram $b_0\oplus S_i$ with two parallel sides $S_i$ and $w+S_i$ if $w\cdot n_i>0$, $P_i(w)=\emptyset$ otherwise. Then 
 it is easy to see that $b(x,w)\in T^\dagger_w(\Om)$ if and only if $x+w\in P_i(w) \bc S_i$ for some $i\in J(w)$. Thus
 \be\label{t}T^\dagger_w(\Om)=\{b(x,w):x+w\in P(w) \cap L\}, \quad P(w) =\bigcup_{i\in J(w)} P_i(w)\bc S_i =(b_0\oplus S_w^+)\bc S_w^+.\ee
 It follows that 
 \be\label{tt}\#T^\dagger_w(\Om)=\# (P(w)\cap L).\ee
 In general, $P(w)$ is not convex. However, if one defines 
 \be\label{ow}\Om_w=b_0\oplus \Om=\bigcup_{t\in [0,1]} (tw+\Om),\ee
 then $\Om_w$ is a convex lattice polygon, being the Minkowski sum of two such sets. In fact,
 \be\label{coh}\Om_w=\hbox{\rm conv}\{\Om,w+\Om\}.\ee
 Also $P(w)=\Om_w\bc\Om$, while $\Om\subset\Om_w$. This and \eqref{tt} imply
 \be\label{dag}\#T^\dagger_w(\Om)=\#(\Om_w\cap L)-\#(\Om\cap L).\ee
 The right hand side can be evaluated using Lemma \ref{pickvar} for each term. Note that $\po_w$ comprises $\po\bc S^+_w$, $w+ S^+_w$ and two $w$-bonds joining these two pieces. In view of  \eqref{sp} the result is
 \be\label{td}\#T^\dagger_w(\Om)=\sum_{i\in J(w)}|S_i| w\cdot n_i+|b_0|/|\bar w|=\sum_{i=1}^N|S_i|\langle w\cdot n_i\rangle +\gcd(w), \ee
 where $\langle x \rangle =(x+|x|)/2$ for $x\in \RR$ and $n=n_i$ on $S_i$ is the unit outward normal to $\po$. Here $|b_0|/|\bar w|=\gcd(w)$.

 It remains to evaluate $T^\ddagger_w(\Om)$. If a bond $b=b(x,w)$ terminates in $w+\Om$, or $x+w\in w+\Om$, then $x\in\Om$. This together with \eqref{t3} and \eqref{t} immediately shows that $b\in T^\ddagger_w(\Om)$ if and only if $x+w\in P(w)\bc w+\Om$. Since $P(w)=\Om_w\bc\Om$, 
 \be\label{ddag} \# T^\ddagger_w(\Om)=\# (Q(w)\cap L), \quad Q(w)= \Om_w\bc (\Om\cup (w+\Om)).\ee
 We will show next that for $|w|$ small enough compared to the facets of $\Om$, $Q(w)$ consists of one or two triangles, each having a vertex at one of the two ends of the simple polygonal line $S_w^+$. For example, if $\Om=[0,3]^2$ and $w=(1,1)$, $Q(w)$ consists of the triangle with vertices $(0,3)$, $(1,4)$ and $(1,3)$ and its image under reflection about the $(1,1)$-axis. Any $b\in T^\ddagger_w(\Om)$ intersects two different facets of $\po$ by \eqref{t3}. Let 
 \be\label{del}\dd=\dd(\Om)=\min_{{\substack{1\le i,j\le N \\ v_i\not\in S_j}}}\hbox{dist}(v_i,S_j)\ee
 where $v_i\in\ZZ^2$ are the vertices of $\Om$.
 The shortest line segment with endpoints on non-adjacent facets has length $\dd$. If $|w|<\dd$, $b\in T^\ddagger_w(\Om)$ necessarily intersects two \emph{adjacent} facets, say $S_i$ and $S_{i-1}$ meeting at some vertex $v_i$, with outward normals $n_i$, $n_{i-1}$ (where $n_0=n_N$). Since both endpoints of $b$ are outside $ \Om$, $w\cdot n_i$ and $w\cdot n_{i-1}$ must have opposite signs. Then in case $w\cdot n_i>0$ and $w\cdot n_{i-1}<0$, $x+w$ is in the triangle with vertices $v_i$, $v_i+w$ and the intersection of $S_i$ and $w+S_{i-1}$, which is therefore part of $Q(w)$. If the reverse inequality holds, the triangle with vertices $v_i$, $v_i+w$ and the intersection of $w+S_i$ and $S_{i-1}$ is part of $Q(w)$. Regarding lattice point count, both cases reduce to the triangle with base $b_0$ and sides normal to $n_i$ and $n_{i-1}$:
 \be\label{triang}T(w,n_i,n_{i-1})=\hbox{conv}\{0,w,q\},\quad q\cdot n_i=0, \quad (q-w)\cdot n_{i-1}=0, \quad (w\cdot n_i)(w\cdot n_{i-1})<0.
 \ee 
 In addition, the relative interior of the base $b(v_i,w)$ of the triangle with endpoints $v_i$, $v_i+w$ is also part of $Q(w)$ and contains $\gcd(w)-1$ lattice points. Consequently, if $|w|<\dd$, 
 \be\label{17}\# T^\ddagger_w(\Om)=\sum_{{\substack{1\le i\le N \\ (w\cdot n_i)(w\cdot n_{i-1})<0}}}[\gcd(w)-1+\#T(w,n_i,n_{i-1})].
\ee
 Unfortunately, $T(w,n_i,n_{i-1})$ is not a lattice polygon in general, since $q$ need not have integer coordinates and Lemma \ref{pickvar} does not apply. Instead, we count the lattice points inside the triangle more directly: 
 
 \begin{lemma}\label{trian} Suppose $(w\cdot n_i)(w\cdot n_{i-1})<0$ and let $T=T(w,n_i,n_{i-1})\subset\RR^2$ be the triangle of \eqref{triang}. Let $u\in\ZZ^2$ be such that $\{u,\bar w\}$ is a lattice basis for $\ZZ^2$. Then
$$\#(\mathring T\cap L)=N_T(w,n_i,n_{i-1}),$$
where for $w\in\ZZ^2$ and unit $n$, $m\in\RR^2$ with $(w\cdot n)(w\cdot m)<0$,
 \be\label{gg}N_T(w,n,m)=s\left(q\cdot\bar w^\perp,q\cdot u^\perp,\gcd(w)\right),\quad q=\frac{m\cdot w}{m\cdot n^\perp} n^\perp\ee
 and $s:\RR \times \RR\times\ZZ\to\RR$ is given by
 \be\label{dede}s(\alpha,\beta,k)=
 (1-\lceil\,|\alpha|\,\rceil)(k-1)+
 \sum_{j=1}^{\lceil\,|\alpha|\,\rceil-1} \left(\left\lceil \frac{\beta -k}{|\alpha|}j\right\rceil -\left \lfloor \frac{\beta }{|\alpha|}j\right\rfloor \right) 
 \ee
 with $s(0,\beta,k)=0$.
 \end{lemma}
 
 \begin{proof} Let $n=n_i$, $m=n_{i-1}$. Since in \eqref{triang} $q\cdot n=0$, $q=\lambda n^\perp$ for some $\lambda\in\RR$. Then solving $(q-w)\cdot m=0$ for $\lambda$ 
gives $q$ as in the second of \eqref{gg}.
 Let $\bar w=w/\gcd(w)=(\bar w_1,\bar w_2)$ and suppose $u=(u_1,u_2)\in\ZZ^2$ solves $u\cdot \bar w^\perp=1$, or $\bar w_2 u_1-\bar w_1 u_2=1$. This is solvable by Bezout's Lemma since $\gcd(\bar w_1,\bar w_2)=1$. Then the matrix $A=\hbox{col}(u,\bw)$ has unit determinant $u\cdot \bar w^\perp=1$ and integer entries, hence so does $A^{-1}=\hbox{row}(\bw^\perp,u^\perp)$. As a result $\{u,\bar w\}$ is a lattice basis for $\ZZ^2$, while the linear transformation with matrix $A^{-1}$ is lattice invariant . 
Now $T'=A^{-1} T$ has vertices $0$, $(0,k)\in\ZZ^2$ and $p=(\alpha,\beta)$, where
\be\label{aux}k=\gcd(w),\quad (\alpha,\beta)=(q\cdot\bar w^\perp,q\cdot u^\perp);\ee
in general $p$ is not a lattice point. Suppose for the moment that $\alpha>0$. Then
 $$\mathring T'\cap L =\left\{(x_1,x_2)\in\ZZ^2: 0<x_1<\alpha, \quad \frac{\beta}{\alpha} x_1< x_2< k+\frac{ \beta -k}{\alpha }x_1\right\}.$$
 For $x\in\RR$ let $\lfloor x \rfloor'$ be the greatest integer \emph{strictly less} than $x$ and 
 $\lceil x \rceil'$ the least integer \emph{strictly greater} than $x$. Then the number of lattice points on a segment $\{(x_1,x_2):x_1=j,\; \mu<x_2<\nu\}$, where $j\in\ZZ$ and $\mu<\nu\in\RR$ equals 
 $\lfloor \nu \rfloor'-\lceil \mu \rceil'+1$. Hence, 
 $$\#(\mathring T'\cap L)= \sum_{j=1}^{\lfloor \alpha \rfloor'} 
 \left(\left\lfloor k+\frac{\beta -k}{\alpha}j\right \rfloor' -\left \lceil \frac{\beta }{\alpha}j\right\rceil'+1\right).$$
 Since $\lfloor x \rfloor'=\lceil x \rceil-1$ and $\lceil x \rceil'=\lfloor x \rfloor+1$, the above reduces to $s(\alpha,\beta,k)$ in \eqref{dede}. It then follows from \eqref{aux} and \eqref{gg} that $\#(\mathring T'\cap L)=N_T(w,n,m)$. The linear transformation with matrix $A$ is lattice invariant and thus $\#(A\mathring T'\cap L) =\#(\mathring T'\cap L)$ \cite{Barpom}, while $AT'=T$. In case $\alpha<0$, reflect $T'$ by replacing $\alpha$ by $|\alpha|$. If $\alpha=0$ then $\mathring T'=\emptyset$.
 \end{proof}
This together with \eqref{17} gives
\be\label{ttt3} \# T^\ddagger_w(\Om)=\sum_{{\substack{1\le i\le N \\ (w\cdot n_i)(w\cdot n_{i-1})<0}}}[\gcd(w)-1+N_T(w,n_i,n_{i-1})].
\ee
To obtain an expression for the $w$-bond number of $\Om$, merely substitute \eqref{pick1}, \eqref{td} and \eqref{ttt3} into \eqref{t4} and rearrange. Observe that for a given bond vector $w$, $N_w(\Om)$ is completely determined by the area $|\Om|$, the lengths $|S_i|$ of the facets, and their orientations through the Miller normals $\bn_i$:

\begin{lemma}\label{bn} Suppose $|w| <\dd$, cf. \eqref{del}. Then the $w$-bond number of $\Om$ is given by
\be\label{bne}\begin{split}N_w(\Om)= |\Om|+\frac{1}{2}\sum_{i=1}^N \frac{ |S_i|}{|\bn_i|}\left(1 -2\langle w\cdot \bn_i\rangle\right)
+1-\gcd(w)\\
+
\sum_{{\substack{1\le i\le N \\ (w\cdot n_i)(w\cdot n_{i-1})<0}}} \!\!\! [\gcd(w)-1+N_T(w,n_i,n_{i-1})].
\end{split}
\ee
 \end{lemma}

\begin{remark}\label{rem2}The above can readily be written in a form similar to \eqref{cpick}:
$$N_w(\Om)=\int_\Om 1 dx+\int_\po g(w,\bn)ds+\sum_{i=1}^N h(w,n_i,n_{i-1}),\quad w\in L,$$ 
as a bulk integral, plus a ``surface'' integral, plus corner contributions, for suitable normal-dependent densities $g$ and $h$.   See also Remark \ref{rem1}.\end{remark}
The present approach of counting bonds has certain similarities with the bond density lemma of Shapeev \cite{Shapeev}. 
 
 \section{Surface Energy of Lattice Polygons}\label{sec3}

We are now in a position to compute the energy. Consider first a \emph{finite-range potential} that only involves bonds within a bounded set. Let \emph{the bond range} $R\subset \Lo$ be symmetric, so that $w\in R\implies -w\in R$. Allow the interatomic potential $\ph_w(\cdot)$ to depend explicitly on $w$, require 
\be\label{fw}\ph_w(\cdot)=\ph_{-w}(\cdot)\; \forall w\in L,\quad \ph_w(\cdot)\equiv 0\; \forall w\in L\bc R,\ee
and define the energy of the homogeneous deformation $y(x)=Fx$, $x\in\Om$, 
\be\label{ey}E\{\Om,y\}=\frac{1}{2}\sum_{x\in\Om\cap L} \sum_{\substack{w\in R\\x+w\in \Om\cap L}}\ph_w(|Fw|),
\ee
where $\ph_w:(0,\infty)\to\RR$ is not restricted to be regular in any way.

\begin{proposition}\label{finener} For $F\in M^{2\times 2}_+$, $\bm\in\ZZ^2$ and unit $n,m\in\RR^2$ define the \emph{stored energy function}
\be\label{w}W(F)=\frac{1}{2}\sum_{w\in R}\ph_w(|Fw|),\ee
the \emph{surface energy density function}
\be\label{se}\gamma_\diamond(F,\bm)=-\frac{1}{4}\sum_{w\in R}\frac{1}{|\bm|} \left( |w\cdot \bm|-1\right)\ph_w(|Fw|)\ee
and the \emph{vertex energy function}
\be\label{ve}\begin{split}&\tau(F,n,m)=\\
&\frac{1}{2} \sum_{w\in R} \left\{\left[H_{n,m}(w)-\frac{1}{2\pi}\theta (n,m)\right] \! (\gcd(w)-1)
+H_{n,m}(w)N_T(w,n,m)\right\}\ph_w(|Fw|),\end{split}\ee
where the sector step function
$$H_{n,m}(w)=\begin{cases} 1 &\mbox{\rm if } (w\cdot n)(w\cdot m)< 0, \\ 
0 & \mbox{\rm if }(w\cdot n)(w\cdot m)\ge 0, \end{cases} $$
and $\theta (n,m)$ is the angle between $n$ and $m$, while $N_T$ is defined in Lemma \ref{trian}.
Suppose the bond range $R$ is bounded with $\max_{w\in R}|w|<\dd$, cf. \eqref{del}. Let $\bn=\bn_i$ on $S_i$. Then the following expression is exact:
\be\label{finen}E\{\Om,y\}=\int_\Om W(F)dx+\int_\po \gamma_\diamond(F,\bn)ds+\sum_{i=1}^N \tau(F,n_i,n_{i-1}).
\ee 
\end{proposition}

 \begin{proof} As in the argument leading to \eqref{eey}, one can write \eqref{ey} as 
 $$ E\{\Om,y\}=
\frac{1}{2}\sum_{w\in R}N_w( \Om)\ph_w(|Fw|).
$$
By the hypothesis on $R$, Lemma \ref{bn} holds for all $w\in R$. Multiply \eqref{bne} by $\ph_w(|Fw|)$ and sum the result over $w\in R$. Interchange the order of summations, noting that $$\sum_{w\in R}\langle w\cdot \bn_i\rangle\ph_w(|Fw|)=\frac{1}{2}\sum_{w\in R}|w\cdot \bn_i|\ph_w(|Fw|)$$ by the symmetry of $R$ and the first of \eqref{fw}, also that the sum of the (dihedral) angles between normals of facets meeting at vertices $\sum_{i=1}^N \theta(n_i,n_{i-1})=1$, and finally that summation over $w$ in the sector of $R$ where $(w\cdot n)(w\cdot m)< 0 $ can be replaced by summation over $R$ provided the summand is multiplied by 
$H_{n,m}(w)$. \end{proof} 

\begin{remark}\label{rem3}  The above result is  not asymptotic  but  \emph{exact}, since we have made no use of asymptotics so far.  It applies to  convex lattice polygons  that are arbitrary apart from the restriction that the bond range is smaller than the characteristic size $\delta$ of  \eqref{del}. \end{remark}

Next we consider \emph{infinite-range potentials}, where $R=\Lo$. We seek the energy of the $k$th dilation $k\Om$ of the region $\Om$, $k\in\ZZ_+$. Here we have no choice but to let $k$ be an integer; otherwise $k\Om$ is not a lattice polygon in general. The following will be useful.
\begin{lemma}\label{corner} Let $M$ be a positive integer. For $\rho>0$ sufficiently large and $p>0$,
 $$\sum_{w\in \ZZ^M,\, |w|>\rho} |w|^{-(M+p)}<C\rho^{-p}$$
 (where $C>0$ is independent of $\rho$ and $p$).
 \end{lemma}
 \begin{proof} Let $\hat x:\RR^M\to \ZZ^M$ be the map $\hat x(\sum_{i=1}^Mx_i e_i)=\sum_{i=1}^M\lfloor x_i\rfloor e_i$, $x_i\in\RR$, with $e_i$ standard basis vectors for $\RR^M$. Thus $|x-\hat x(x)|\le D$, the unit cell diameter. Write $\hat x(x)=x+(\hat x(x)-x)$ and invoke the triangle inequality to conclude $|x|-D\le |\hat x(x)|\le |x|+D$. Then also $|\hat x(x)|\ge \rho$ implies $|x|>\rho-D$, while $|\hat x(x)|^{-(M+p)}\le| |x|-D|^{-(M+p)}$. Thus $A_\rho=\{x\in\RR^M:|\hat x(x)|\ge \rho\}\subset \RR^M \bc B_{\rho-D}(0)$.
 \begin{gather*}0<\sum_{w\in \ZZ^M \bc B_\rho(0)} |w|^{-(M+p)}= \int_{A_\rho} |\hat x(x)|^{-(M+p)}dx\le C \int_{\rho-D}^\infty (r-D)^{-(M+p)}r^{M-1}dr
 \\ \le C \int_{\alpha \rho}^\infty (\alpha r)^{-(M+p)}r^{M-1}dr 
 =C \int_{\rho}^\infty r^{-(1+p)}dr=C\rho^{-p},\end{gather*}
 where $\alpha\in (0,1)$ is such that $\alpha \rho=\rho-D$, so that $\alpha\in (1/2,1)$ for $\rho>2D$ and $C$ is a generic constant with possibly different values each time it appears.
 \end{proof} 

For convenience we suppose that the interatomic potential $\ph_w(\cdot)=\ph(\cdot)$ (does not explicitly depend on $w$), although this is not essential.
\begin{proposition}\label{infener} Suppose the interatomic potential $\ph:(0,\infty)\to\RR$ satisfies the following: for each $r_0>0$ and for some constants $C=C(r_0)$ and $d>2$, 
\be\label{ph} |\ph(r)|<Cr^{-(2+d)} \quad \hbox{for }r\in [r_0,\infty).\ee
Let the bond range $R=\Lo$ in \eqref{ey} and in the definitions \eqref{w} of $W$, \eqref{se} of $\gamma_\diamond$ and \eqref{ve} of $\tau$. Then as $k\to\infty$, $k\in\ZZ_+$,
\be\label{infen}E\{k\Om,y\}=k^2\int_\Om W(F)dx+k\int_\po \gamma_\diamond(F,\bn)ds+\sum_{i=1}^N \tau(F,n_i,n_{i-1})+O(k^{2-d}).
\ee
\end{proposition}
\begin{proof} Note that $\dd(k\Om)=k\dd(\Om)=k\dd$ in \eqref{del}, so that Lemma \ref{bn} for $k\Om$ holds provided 
 \be\label{rk}w\in R_k= (\Lo)\cap B_{k\dd}(0)=(L\cap B_{k\dd}(0))\bc \{0\}.\ee
 Split the energy as follows:
\be\label{split} E\{k\Om,y\}=
\frac{1}{2}\sum_{w\in R_k}N_w(k \Om)\ph(|Fw|)+\frac{1}{2}\sum_{w\in L\bc R_k}N_w( k\Om)\ph(|Fw|).
\ee
Now it is clear that for any $w\in L$ and $k\in\ZZ_+$,
$$0\le N_w(k\Om)\le\#(k\Om\cap L)<Ck^2$$
for some constant $C>0$, since all bonds within $k\Om$ start in $k\Om$ and by Lemma \ref{pickvar} applied to $k\Om$ (the dominant term in \eqref{pick1} would be $|k\Om|=k^2\Om$).
This provides a bound for the second term in \eqref{split}:
\be\label{est1}
\begin{split}
\left| \sum_{w\in L\bc R_k}N_w( k\Om)\ph(|Fw|)\right| <& Ck^2\sum_{w\in L\bc R_k} |\ph(|Fw|)|\\ < Ck^2\!\!\!\sum_{w\in \ZZ^2,\, |w|>k\dd} |\alpha w|^{-(2+d)}<Ck^{2-d},
\end{split}
\ee
where we invoked \eqref{ph}, $\alpha>0$ is such that $|Fz|>\alpha |z| $ for all $z\in\RR^2$ and we used Lemma \ref{corner} with $\rho=k\dd$ and $p=d$; and $C$ is a generic constant with possibly different values each time it appears. 

The first term in \eqref{split} is covered by Proposition \ref{finener} applied to $k\Om$, since $w\in R_k$ means $|w|<k\dd=\dd(k\Om)$.
Noting that $|k\Om|=k^2|\Om|$, $|k S_i|=k|S_i|$, Proposition \ref{finener} implies
\be\label{bru}\frac{1}{2}\sum_{w\in R_k}N_w(k \Om)\ph(|Fw|)=k^2|\Om| W_k(F)+k\sum_{i=1}^N |S_i| \gamma_k(F,\bn_i)+\sum_{i=1}^N \tau_k(F,n_i,n_{i-1}),\ee
where $W_k$, $\gamma_k$ and $\tau_k$ are given by \eqref{w}, \eqref{se} and \eqref{ve} with $R_k$ in place of $R$; see \eqref{rk}. Recalling that $W$, $\gamma_\diamond$ and $\tau$ are defined by the same equations with $R=\Lo$, using Lemma \ref{corner} with $M=2$, we may estimate (omitting arguments)
\be\label{est3}|W-W_k|<Ck^{-d},\quad |\gamma-\gamma_k|<Ck^{1-d},\quad |\tau -\tau_k|<Ck^{2-d}.\ee
We only demonstrate the third of these, the others being easier. Recall that in \eqref{ve}, $N_T$ is the number of lattice points in the interior of a certain triangle $T$ whose area is bounded above by $C|w|^2$, cf. Lemma \ref{trian}. By Pick's Theorem \eqref{pickk} (applied to the lattice parallelogram of smallest area $A$ containing $T$, and having the same base) the area $A$ exceeds $N_T$ hence $N_T<C|w|^2$. Also $\gcd(w)\le|w|$, $|H_{n,m}|\le 1$, hence we have from \eqref{ve}, 
$$ |\tau -\tau_k|< \sum_{w\in \ZZ^2,\, |w|>k\dd} C |w|^2|\ph(|Fw|)|< \,\, C\!\!\!\!\sum_{w\in \ZZ^2,\, |w|>k\dd}|w|^{-d}<Ck^{2-d}$$
proceeding as in \eqref{est1}. By \eqref{est3}, replacing $W$, $\gamma_\diamond$ and $\tau$ by $W_k$, $\gamma_k$ and $\tau_k$ in \eqref{bru} produces an error of $O(k^{2-d})$. Combine this with \eqref{est1} and \eqref{split} to obtain \eqref{infen}.
\end{proof} 

\section{Surface Energy for More General Boundaries}\label{sec4}

 We examine the surface energy density function $\gamma_\diamond$ in \eqref{se} more closely, paying attention to its dependence on the surface normal. Due to its construction, $\gamma_\diamond(F,\cdot)\colon \bbm\to\RR$ is defined only for ``rational directions'', that is, on the set of Miller normals
 \be\label{miller2}\bbm=\{\bn:\bn=(\nu_1,\nu_2)\in\ZZ^2,\; \gcd(\nu_1,\nu_2)=1\}.\ee
 Using \eqref{w}, we rewrite $\gamma_\diamond$ in \eqref{se} as
 \be\label{see}\gamma_\diamond(F,\bn)= -\frac{1}{4}\sum_{w\in R} \left|w\cdot n\right|\ph(|Fw|)\;+\frac{1}{2|\bn|}W(F), \quad n=\bn/|\bn|,\quad \bn\in \bbm.\ee
 The first term (involving the sum) reduces to a function of the unit normal $n$, and trivially admits a unique continuous extension onto the whole of the unit circle $S^1$. There is no such extension for the second term. Define the \emph{rational} and \emph{irrational direction sets} as
 \be
 \label{rir}S^1_R=\{n:n\in S^1,\; n=\bn/|\bn|,\; \bn\in \bbm\},\quad S^1_I=S^1\bc S^1_R,
 \ee
respectively, where $ \bbm$ is defined in \eqref{miller2}. Thus a vector is rational (irrational) if the tangent of the angle it makes with the usual basis vectors is rational (irrational). Since facets of lattice polygons have rational normals, the surface energy density $\gamma_\diamond$ is defined only for such directions. Note that for each $n\in S^1_R$ there is a unique $\bn=\bn(n)\in \bbm$ with $\bn/|\bn|=n$. 
The question arises as to how one can extend the definition of $\tilde\gamma_\diamond(F,n)=\gamma_\diamond(F,|\bn(n)|n)$, $n\in S^1_R$, to the whole of $S^1$. This is related to another question: what is the surface energy when $\po$ is \emph{smooth}, for example $\po=S^1$? It turns out that this question can be answered, at least partially, using the present approach. The basic idea is that even if $\po$ is not polygonal, but smooth, the convex hull of all lattice points inside $\Om$ is a convex lattice polygon. 

\begin{proposition}\label{smopo} Let $\Om\subset\RR^2$ be strictly convex and $\po$ be $C^2$ with positive curvature. Suppose $\ph$ is as in Proposition \ref{infener}, but with $d>3$. Define the reduced surface energy density $\gamma_\circ:M^{2\times 2}_+\times S^1\to\RR$ by
\be\label{go}
\gamma_\circ(F,m)= -\frac{1}{4}\sum_{w\in \Lo} \left|w\cdot m\right|\ph(|Fw|), \quad F\in M^{2\times 2}_+, \quad m\in S^1.
\ee
Then for any sequence $r=r_k\to\infty$ as $k\to\infty$ ($r_k\in\RR_+$, $k\in\ZZ_+$),
\be\label{zazen}E\{r\Om,y\}=r^2\int_\Om W(F) dx +r\int_\po \gamma_\circ(F,n)ds+O(r^{2/3}),
\ee
where $n:\po\to S^1$ is the unit outward normal to $\po$.
\end{proposition}

\begin{remark}\label{rem4} This  asymptotic  result for inflated regions $r\Om$, is \emph{sequence-independent};  that is, the sequence of $r\to\infty$ is not restricted to be integer but is arbitrary. This  occurs because the lattice point remainder  $R(  r)=O(r^{2/3})$ \cite{C,huxley} is  of lower order than the surface energy.  In contrast, the surface energy for lattice polygons, or the more general regions considered in Proposition \ref{conpo}  depends on the sequence of dilation factors. The dependence on the dilation sequence is thoroughly studied in one dimension  in \cite{Mora}.\end{remark}

\begin{proof} 
For each $r>0$, let $\Om_r=\hbox{conv}(r\Om\cap L)$. Then $\Om_r\subset r\Om$ is a convex lattice polygon, while $r\Om\cap L =\Om_r\cap L$. Hence, in view of \eqref{eeenergyy},
\be\label{eq}E\{r\Om,y\}=E\{\Om_r,y\},\ee
where $y(x)=Fx$ for $x\in r\Om$. The calculation of $E\{\Om_r,y\}$ proceeds as above with one exception. For any $w\in L$, the condition $|w|<\delta(\Om_r)$ (see \eqref{del}) may be violated for large enough $r$, since facets may become as small as the shortest lattice bonds. This affects $N_w(\Om_r)$, but only the part regarding $ \# T^\ddagger_w(\Om_r)$---see \eqref{t4}, \eqref{ddag}---which we merely need to estimate. Given any convex body $D\subset\RR^2$, let
$$\hat Q_w(D)=(b_0\oplus D)\bc [D\cup( w+D)],\quad Q_w(D)=\hat Q_w(D)\cap L,$$
where $b_0=\hbox{conv}\{0,w\}$.
Then by \eqref{ddag} and \eqref{coh}, $ \# T^\ddagger_w(\Om_r)=\# Q(\Om_r)$.
 Since $\Om_r\subset r\Om$, and $r\Om\bc\Om_r$ contains no lattice points, \eqref{ddag} and \eqref{coh} imply
 \be\label{eee1} Q_w(\Om_r)\subset Q_w(r\Om).\ee
Let $q$, $q'\subset \p(r\Om)$ be the two points of $\p(r\Om)$ where the tangent vector is $w$, and $B_{r\rho}\subset r\Om$ be a disk with $\p B_{r\rho}$ tangent to
$\p(r\Om)$ at $q$, where $\rho$ is the smallest radius of curvature of $\po$. Also let $B'_{r\rho}\subset r\Om$ be a similar disk tangent to $\p(r\Om)$ at $q'$.
Then for $r$ large enough it is easy to see that 
\be\label{eee2}Q_w(r\Om)\subset Q_w(B_{r\rho})\cup Q_w(B'_{r\rho}).\ee
The connected component of $\hat Q_w(B_{r\rho})$ containing $q$ is contained inside an isosceles triangle with base a $w$-bond (with length $|w|$), and height the distance from the base middle to the intersection of the two circles $\p B_{r\rho}$ and $w+\p B_{r\rho}$; these are tangent to the base at its endpoints. The triangle height is thus bounded by $C(r)|w|$, where $C(r)$ approaches zero for large $r$. A crude but sufficient upper bound of the lattice point count of this set, hence also of the right hand of \eqref{eee2}, is $C|w|^2$, with $C$ independent of $r$. In view of of \eqref{eee1}, $\# T^\ddagger_w(\Om_r)$ is also bounded by $C|w|^2$. This estimate replaces the sum over vertices (second sum) in \eqref{bne}. Since $\sum_{w\in \Lo}|w|^p\ph(|Fw|)$ are absolutely convergent for $p=0,1,2$ as one infers from Lemma \ref{corner}, it follows that
\begin{align}
\label{frog}
E\{\Om_r,y\}=&|\Om_r|W(F)+ \int_{\p\Om_r} \gamma_\diamond(F,\bn)ds+O(1)\nonumber\\
=&\left[|\Om_r|+ \int_{\p\Om_r}\frac{1}{2|\bn|}ds\right]W(F)+ \int_{\p\Om_r} \gamma_\circ(F,n)ds+O(1)\nonumber\\
=&\#\left( r\Om\cap L \right) W(F)+ \int_{\p\Om_r} \gamma_\circ(F,n)ds+O(1).
\end{align}
Here we have used \eqref{see} and \eqref{go}, then \eqref{cpick}, in which the last term (sum) equals $1$, together with the fact $r\Om\cap L =\Om_r\cap L$.
We turn to $\int_{\p\Om_r} \gamma_\circ(F,n)ds$. Recalling \eqref{go}, a typical term involves 
\be\label{55}\int_{\p\Om_r} |w\cdot n|ds=2|\hbox{Proj}_{w^\perp} \po_r||w|,\ee
$|\hbox{Proj}_{w^\perp} \po_r|$ being the length of the projection of $\po_r$ onto a line perpendicular to $w$. This follows after splitting $\po_r$ into two pieces, over which $w\cdot n$ is $\ge0$ and $\le0$, and using the Divergence Theorem on each.
Next, we show that 
\be\label{seg}0<|\hbox{Proj}_{w^\perp} \p(r\Om)|-|\hbox{Proj}_{w^\perp} \po_r| <C|w|^2,
\ee
where the constant $C$ is independent of $r>1$ and $w$. There are lattice points $z^-$ and $z^+\in\p\Om_r\cap L$, such that $\Om_r$ lies entirely between lattice lines $l^-$, $l^+$ with normal $w^\perp$ and containing $z^-$, $z^+$, respectively. Consider the part of $\p(r\Om)$ that lies outside the strip bounded by $l^+$ and $l^-$. It consists of two disjoint arcs, one to the ``right'' of $l^+$ and the other to the ``left'' of $l^-$. The length of the projections of these two arcs onto the $w^\perp$ axis equals the difference in \eqref{seg}. Let $c^+$ be the arc to the right of $l^+$ (with endpoints in $l^+$). Let $s$ be the region bounded by $c^+$ and $l^+$. The only lattice points it contains are in $l^+$. This is true since $r\Om\bc\Om_r$ is free of lattice points. By the strict convexity of $r\Om$, there is a unique $q\in c^+$ where the normal to $c^+$ is $w^\perp$. Consider the osculating circle of $c^+$ at $q$. Let $s'\subset s$ be the portion of the osculating disc contained in $s$; it is a circular segment whose height (in the direction $w^\perp$) equals the thickness of $s$ (the length of its projection onto a line along $w^\perp$). The radius of the circle is $r\rho$ for some $\rho>0$. There are two possibilities. Either $s'$ lies between $l^+$ and the next lattice line $l'$ with normal $w^\perp$ to the right of $l^+$, or it extends beyond $l'$ to the right. In the first case the height of the segment $s'$ is $1/|\bw|$, the distance between adjacent lattice lines with normal $w^\perp$. In the second case, let $s''$ be the portion of $s'$ to the right of $l'$. Then $s''$ is also a circular segment and free of lattice points. Suppose its chord length is $c$ and height is $h$. Since the radius of the circular arc is $r\rho$, we have $h^2-2r\rho h +c^2/4=0$. Solving this for $h/(r\rho)$ and using the inequality $1-\sqrt{1-x}<x$ for $0<x<1$ yields $h< c^2/(4r\rho)$. Now since the circular segment $s''$ is free of lattice points and its chord is in $l'$, the chord length $c<|\bw|\le|w|$ (since the distance between adjacent lattice points in $l'$ is $|\bw|$.) Hence $h<|w|^2/(4r\rho)$. The total height of the larger circular segment $s'$ is $h+1/|\bw|$ which is thus bounded by $C|w|^2$ for $r\ge1$. The thickness of $s$ in the direction normal to $w$ is the same as this height. This shows \eqref{seg}.

Combining \eqref{seg} with \eqref{55} shows 
\be
\label{ww3}
\left|\int_{\p(r\Om)} |w\cdot n|ds-\int_{\p\Om_r} |w\cdot n|ds\right|<C|w|^3.
\ee
In view of \eqref{go} and since the sum $\sum_{w\in \Lo}|w|^3\ph(|Fw|)$ converges absolutely by hypothesis, one deduces
\be\label{56}\int_{\p\Om_r} \gamma_\circ(F,n)ds= \int_{\p(r\Om)} \gamma_\circ(F,n)ds+O(1)=r\int_{\po} \gamma_\circ(F,n)ds+O(1).\ee
Our hypotheses on $\po$ ensure that $\#(r\Om\cap L)=r^2|\Om|+O( r^{2/3})$, e.g. \cite{huxley,ivic}. This together with \eqref{56} in \eqref{frog} and \eqref{eq} proves \eqref{zazen}.
 \end{proof}

According to Proposition \ref{smopo}, when $\po$ is smooth and strictly convex, so that the normal vector is \emph{irrational} almost everywhere on $\po$, the surface energy density is given by \eqref{go}; in contrast, for lattice polygons (with rational normal a.e. on $\po$), the surface energy density is given by \eqref{see}. This suggests that we combine the two expressions in defining a surface energy density for all values of the unit normal.
That will allow us to treat a more  general case with $\Om$  a (not necessarily strictly) convex body. We do place some restrictions on $\po$: flat parts of $\po$ must be lattice segments (with rational normals). Corners have to be lattice points.

We will need the following auxiliary result:

\begin{lemma}\label{lem1/n}Let $D\subset\RR^2$ be a strictly convex body and $\pd$ be $C^2$ with positive curvature. For $r>0$ define the convex lattice polygon $D_r=\hbox{conv}(r D\cap L)$ with Miller normal $\bn:\pd_r\to \bbm$. Then as $r\to\infty$,
$$ \int_{\pd_r}\frac{1}{|\bn|}ds=O(r^{2/3}).$$
\end{lemma}

\begin{proof} By Pick's Theorem (Lemma \ref{pickvar}), and since $\#(D_r\cap L)=\#(rD\cap L)$,
$$ \int_{\pd_r}\frac{1}{2|\bn|}ds=\#(D_r\cap L)-|D_r|-1=\#(rD\cap L)- |r D|+ |r D|-|D_r|-1.$$
Now $\#(rD\cap L)- |r D|=o( r^{2/3})$ 
 by \cite{C,huxley}. In view of Theorem 4 and Remark 2 of \cite{barany}, and since $D_r\subset r D$,
\be
\label{barany}
0< |r D|-|D_r|<C r^{2/3}
\ee
for some constant $C$. The result follows.
\end{proof}

We now state the main result of this section:

\begin{proposition}\label{conpo} Assume that $\Om$ is a convex body with $\po$ Lipschitz, and that there is a finite set of lattice points $\{v_1,\ldots,v_N\}\subset \po\cap L$, that partitions $\po$ into $N$ curves $S_i$, $ \po=\bigcup_{i=1}^N S_i$, each with endpoints $v_i$ and $v_{i+1}$ ($v_{N+1}=v_1$), such that $S_i\cap S_{i+1}=v_{i+1}$, $S_i$ is a $C^2$ curve and one of the following two alternatives holds: \\
(i) For $i\in J_c\subset\{1,\ldots,N\}$, $S_i\subset \GG_i$, where $\GG_i$ is a simple closed $C^2$ curve with positive curvature, or \\
(ii) For $i\in J_f=\{1,\ldots,N\}\bc J_c$, $S_i$ is a straight segment. \\
Suppose $\ph$ is as in Proposition \ref{infener}, but with $d>3$. Define
the \emph{extended surface energy density} $\hgg(F,\cdot)\colon S^1\to\RR$ as follows:
\be\label{gh}\hgg(F,n)
=\begin{cases} 
\displaystyle{ -\frac{1}{4}\sum_{w\in \Lo}
 \left |w\cdot n\right|\ph(|Fw|)
+\frac{1}{2|\bn|}W(F)},
& n\in S^1_R, \;\bn\in \bbm,\;\bn/|\bn|=n,
\\ & 
\\ \displaystyle{ -\frac{1}{4}\sum_{w\in \Lo}
 \left |w\cdot n\right|\ph(|Fw|)}=\gamma_\circ(F,n),&n\in S^1_I,
\end{cases}
\ee
with $\gamma_\circ$ defined in \eqref{go} and $S^1_R$, $S^1_I$ defined in \eqref{rir}.
 Then as $ k\to\infty$, $k\in\ZZ_+$,
\be
\label{muzen}E\{k\Om,y\}=k^2\int_\Om W(F) dx +k\int_\po \hgg(F,n)ds+O(k^{2/3}),
\ee
where $n:\po\to S^1$ is the unit outward normal to $\po$.
\end{proposition}

\begin{proof} We now choose $r=k\in\ZZ_+$ and let $\Om_k=\hbox{conv}(k\Om\cap L)$. The part of the proof of Proposition \ref{smopo} prior to \eqref{frog} is easily adapted to the present setting, so that once again, as $k\to\infty$, with $\gamma_\diamond$ as in \eqref{see},
\be
\label{ffrog}
E\{\Om_k,y\}=|\Om_k|W(F)+ \int_{\p\Om_k} \gamma_\diamond(F,\bn)ds+O(1).
\ee
Let $\po_f$ be the union of those $S_i$ that are straight segments and $\po_c$ the union of the $S_i$ with positive curvature, so that $\po=\po_f\cap \po_c$. By hypothesis, for $k\in\ZZ_+$ we have $kv_i\in\p(k\Om)\cap L$, hence also 
$kv_i\in\p(\Om_k)\cap L$. Then
$k\po_f\subset\po_k=\p(\Om_k)$. Let $\po_k^c=\po_k\bc k\po_f$. Then 
$$
E\{\Om_k,y\}=|\Om_k|W(F)+ \int_{\po_k^c} \gamma_\diamond(F,\bn)ds+ \int_{k\po_f} \gamma_\diamond(F,\bn)ds+O(1).
$$
Our hypotheses regarding $\po_c$, specifically alternative (i), ensure that $n\in S^1_I$ a.e. on $k\po_c$, while (ii) implies that $n\in S^1_R$ a.e. on $k\po_f$. Using \eqref{gh}, rewrite the above as
\be\label{unmon}
E\{\Om_k,y\}= |k\Om|W(F)+ \int_{k\po_c} \gamma_\circ(F,n)ds+ \int_{k\po_f} \hgg(F,n)ds+{\hat R}(k),
\ee
where
\be
\label{rkk}
\begin{split}
{\hat R}(k)=\left[|\Om_k|-  |k\Om| + \int_{\po_k^c}\frac{1}{2|\bn|}ds \right] & W(F) \\ +  \int_{\po_k^c} \gamma_\circ(F,n)ds-  & \int_{k\po_c} \gamma_\circ(F,n)ds+O(1).
\end{split}\ee
It remains to show that ${\hat R}(k)=O(k^{2/3})$ as $k\to\infty$, $k\in\ZZ_+$. Let $i\in J_c$, so that $S_i$ satisfies alternative (i) in the statement of Proposition \ref{conpo}. Let $S_k^i$ be the portion of $\po_k^c$ between $kv_i$ and $kv_{i+1}$, i.e., terminating at these two points and containing no other $kv_j$.
Let the strictly convex body $D^i$ be such that $\p D^i=\GG_i$. Let $G_k^i$ be the bounded region whose boundary is $kS_i\cup S_k^i$; this is well defined since both curves terminate at $kv_i$ and $kv_{i+1}$. Then $G_k^i\subset kD^i\bc D_k^i$, where $D_k^i=\hbox{conv}(k D_i\cap L)$, and
\be
\label{zz1}
|k\Om\bc \Om_k|=\sum_{i\in J_c}|G_k^i|\le \sum_{i\in J_c}| kD^i\bc D_k^i| <C k^{2/3}
\ee
in view of \eqref{barany} applied to $D^i$ for $r=k\in\ZZ_+$.

Next, note that $S_k^i\subset \p D_k^i$. As a result, 
\be
\label{zz2}
0< \int_{\po_k^c}\frac{1}{2|\bn|}ds= \sum_{i\in J_c}\int_{S_k^i}\frac{1}{2|\bn|}ds \le 
 \sum_{i\in J_c}\int_{\p D_k^i}\frac{1}{2|\bn|}ds <C k^{2/3}
\ee
by Lemma \ref{lem1/n} with $D=D^i$.

Next, we turn to the difference of the last two integrals in \eqref{rkk}. Recalling \eqref{go}, we write this as follows:
$$
\sum_{i\in J_c} \sum_{w\in \Lo}\ph(Fw) I_k^i(w), \quad I_k^i(w)=\int_{kS_i} |w\cdot n| ds
-\int_{S_k^i} |w\cdot n| ds
=\int_{\p G_k^i} |w\cdot \tilde n| ds,
$$
where $n$ is the outward unit normal to $k\po$ and $\po_k$ in the first two integrals, while $\tilde n$ is outward unit normal to $\p G_k^i$. Hence $ I_k^i(w)>0$, and since $G_k^i\subset kD^i\bc D_k^i$, 
$$0< I_k^i(w) \le \int_{\p(k D^i)} |w\cdot n| ds
-\int_{\p D_k^i} |w\cdot n| ds <C|w|^3,
$$
where the estimate follows from \eqref{ww3} by replacing $\Om$ of Proposition \ref{smopo} by $D^i$; the constant $C$ is independent of $k$.
Since the sum $\sum_{w\in \Lo}|w|^3\ph(|Fw|)$ converges absolutely by hypothesis, so does the double sum in the previous equation; therefore
$$
\int_{\po_k^c} \gamma_\circ(F,n)ds- \int_{k\po_c} \gamma_\circ(F,n)ds=O(1).
$$
This together with \eqref{zz1} and \eqref{zz2} shows that ${\hat R}(k)=O(k^{2/3})$.
 The normal is irrational a.e. on $\po_c$. Consequently $\int_{k\po_c}\gamma_\circ ds=\int_{k\po_c}\hgg ds=k\int_{\po_c}\hgg ds$, and \eqref{muzen} follows from \eqref{unmon}, since \eqref{eq} holds.
\end{proof}

\begin{remark}\label{rem9}It is interesting that in cases where the normal is rational on a subset of $\po$ of positive measure, the dilation factors are required to be integers. In contrast, the result of Proposition \ref{smopo} (where the normal is  irrational  almost everywhere on $\po$) is  independent of the sequence of dilation factors. In one dimension it is known \cite{Mora} that the coefficients in the asymptotic expansion of the energy depend on this sequence. It should be kept in mind that there is no counterpart in one dimension of an irrational surface, which is purely a higher-dimensional occurrence. The reason for the difference between the rational and irrational cases is the different  order of the lattice point remainder term.\end{remark}

Proofs of the Wulff theorem associated with surface energy minimization \cite{fonseca} over domains with given measure typically rely on continuity of the surface energy density with respect to the unit normal; see \cite{dac} and Remark \ref{rem6} for a weaker alternative. Perhaps surprisingly, the extended surface energy density $\hgg(F,\cdot)\colon S^1\to\RR$ exhibits a dense set of discontinuities as we show next.

\begin{proposition}\label{propg} Suppose $\ph$ is as in Proposition \ref{infener} and fix $F\in M^{2\times2}_+$. Then 
\item (i) $\gamma_\circ(F,\cdot)\colon S^1\to\RR$ in \eqref{go} is Lipschitz continuous on $S^1$.
\item (ii) $\hgg(F,\cdot)\colon S^1\to\RR$ defined in \eqref{gh} is continuous at $n\in S^1_I$, discontinuous at $n\in S^1_R$ and differentiable at most on a subset of $S^1_I$ of measure zero.
\end{proposition}

\begin{proof} Arrange the elements of $\Lo$ in a sequence: $\{w_j\}$, $j=1,2,\ldots,$ such that $|w_{j+1}|\ge|w_j|$, and define 
$g_j(n)=(-1/4)\ph(|F w_j|)|w_j\cdot n| $ for $n\in S^1$. Then clearly $g_j\colon S^1\to \RR$ is Lipschitz on $S^1$ and (formally for the moment)
$\gamma_\circ(F,n)=\sum_{j=1}^\infty g_j(n)$. Now since $|g_j|\le M_j=\left| \ph(|F w_j|)\right| \,|w_j|$ on $S^1$ and the series $\sum_{j=1}^\infty M_j=\sum_{w\in\Lo}\left| \ph(|F w|)\right| \,|w|$ converges in view of Lemma \ref{corner}, then $G_k(n)=\sum_{j=1}^k g_j(n)$ converge uniformly as $k\to\infty$ to $\gamma_\circ(F,n)$ on $S^1$ by the Weierstrass M test.
Since $n\mapsto |w\cdot n|$, $n\in S^1$ is Lipschitz with constant $|w|$, the Lipschitz constant of $G_k$ is bounded above by
$$\sum_{j=1} ^k \left| \ph(|F w_j|)\right| \,|w_j|<\sum_{w\in\Lo} \left| \ph(|F w|)\right| \,|w|<\infty.
$$
The uniform convergence of the $G_k$ together with the uniform bound on their Lipschitz constants guarantee that the limit function $\gamma_\circ(F,\cdot)$ is also Lipschitz on $S^1$ and (i) holds.

To show (ii), consider the function 
$$
 h(n)
=\begin{cases} 
 \frac{1}{|\bn|},
& n\in S^1_R \;\;(\bn\in \bbm,\;\bn/|\bn|=n),
\\ & 
\\ 0,&n\in S^1_I.
\end{cases}
$$
In other words, letting $n=(\nu_1,\nu_2)\in S^1$,
\be\label{hhh} h(\nu_1,\nu_2)
=\begin{cases} 
 \frac{1}{\sqrt{p^2+q^2}},
& \nu_2/\nu_1=p/q,\;\;(p,q)\in\ZZ^2,\;\; \hbox{gcd}(p,q)=1,
\\ 0,&\nu_1=0, 
\\ 0,&\hbox{otherwise.}
\end{cases}
\ee
Then one has
\be\label{ghh}
\hgg(F,n)=\gamma_\circ(F,n)+\frac{1}{2}W(F)h(n)\quad\forall n\in S^1.
\ee
By (i), it suffices to prove that $h$ is continuous at irrational $n$ and discontinuous at rational $n$ to show the continuity part of (ii). In fact, $h$ is very similar to the Thomae function $T(x)=1/q$ for $x=p/q$, $p$, $q$ coprime integers ($x$ rational), and zero for $x$ irrational; see e.g. Proposition 4.1 in \cite{sally}. Adapting these results to  $h$ is trivial in view of \eqref{hhh}. Thus $h$ is continuous at irrational $n$ and discontinuous at rational $n$ and so is $\hgg(F,\cdot)$. Also $h$ is nowhere differentiable by a simple adaptation of Proposition 6.1, \cite{sally}. Since by part (i) $\gamma_\circ$ is Lipschitz, it is differentiable a.e. on $S^1$ by the Rademacher theorem. Then $\hgg(F,\cdot)$ fails a.e. to be differentiable by \eqref{ghh}. Also it is not differentiable at rational $n$ as it is not continuous there. 
\end{proof}

 \section{A Continuous Surface Energy Density}\label{sec5} 
There are two issues associated with the surface energy density  $\hgg$. The first issue  is the lack of continuity of $\hgg(F,\cdot)$. This suggests that the \emph{surface energy minimization problem}, that of  minimizing the integral $\int_\po\hgg(F,n)ds$ over a suitable class of regions $\Om$ with $|\Om|$ fixed, may actually  be ill posed. 

\begin{remark}\label{rem6}The standard hypothesis for surface energy minimization in three dimensions  is continuity of   $\hgg(F,\cdot)$ \cite{fonseca}. However,  in two dimensions, as shown by Dacorogna and Pfister \cite{dac},    lower semicontinuity of $\hgg(F,\cdot)$ suffices. It is  easy to show from \eqref{gh} and Proposition \ref{propg} that  $\hgg(F,\cdot)$ is indeed lower semicontinuous provided $W(F)\le 0$.  The latter inequality is not unreasonable; for example,  it is satisfied for  values of $F$ near the  minimum of $W(F)$,  when the latter is given by  \eqref{0} with $\ph$ a standard Lennard-Jones potential.
\end{remark}
The second issue is that the surface energy minimization problem with density $\hgg(F,\cdot)$ is \emph{not  physically appropriate}, since  fixing $|\Om|$ is not the same as  fixing the total mass, or  equivalently, the number $\#(\Om\cap L)$ of lattice points of $\Om$. If the minimization were over the class of lattice polygons with fixed lattice point number, the appropriate constraint would fix $|\Om|+\int_\po 1/(2|\bn|) ds$ instead of $|\Om|$, by virtue of Lemma  \ref{pickvar}.  
For a lattice polygon, the lattice point remainder $R(k)=\#(k\Om\cap L)-|k\Om|$ can be written as 
\be\label{remainder}R(k)=k\int_{\po}1/(2|\bn|) ds+1,\quad k\in\ZZ.\ee
using Lemma \ref{pickvar}. It seems that  $R$ is implicated in both issues raised  above.  Being $O(k)$, it contributes to the surface energy and gives rise to the term $\frac{1}{2|\bn|}W(F)$ in \eqref{gh}, \eqref{ghh}, which is the one responsible for the lack of continuity of $\hgg$.  Also, surface energy minimization over domains with fixed measure would seem to  make physical sense only if their lattice point remainder $ \#(\Om\cap L)-|\Om|$ vanishes, so that  constraining $|\Om|$  fixes the lattice point number, hence the mass (see also Remark \ref{rem7.5} below).
One way to ensure this might be to seek a sequence of dilation factors $r_k\in\RR$ satisfying condition \eqref{bbl} imposed by \cite{Blanc}, i.e., $R(r_k)=0$.  It is not clear for what choices of $\Om$ this is possible, and we consider two ways to modify this approach.

\emph{First, we relax the condition  $R(r_k)=0$ and require instead that there is a sequence  $r_k$ such that}
\be\label{relseq}R(r_k)=o(r_k),\quad r_k\to \infty,\ee
so that the lattice point remainder is of  lower order  than the surface energy, which is $O(r_k)$. This is satisfied for the smooth regions with positive boundary curvature of Section \ref{sec4}, where the fact that $R(  r)=O(r^{2/3})$ for any real sequence $r\to\infty$ was exploited in proving Proposition \ref{smopo}.  As a result, the density $\gamma_\circ$ in \eqref{zazen}  is  continuous in the unit normal by extension to the whole of $S^1$; see Proposition \ref{propg} (i).

\emph{Second, in case $\Om$ is a lattice polygon, or the ``mixed'' region of Proposition \ref{conpo},  we rewrite the energy in terms of   an ``equivalent'' region $\Om(k)$ containing the same lattice points as the scaled region $k\Om$.  
}  Accordingly, from \eqref{eeenergyy} it is clear that 
$E\{k\Om,y\}=E\{\Om(k),y\}$. Observe that, given the set  of atoms that are within a convex region $k\Om$, there is some freedom in choosing an alternative convex region $\Om(k)$ \emph{containing precisely the same atoms}. By choosing $\Om(k)$ in a specific way, we can ensure that the lattice point remainder of $\Om(k)$ is of lower order than the surface energy. In effect, this is equivalent to \eqref{relseq} apart from the fact that $\Om(k)$ need not be a dilation of $\Om$.  For lattice polygons, $\Om(k)$  can be constructed as follows. The ``interplanar'' distance between adjacent parallel lattice lines with Miller normal $\bn$ is 
$1/|\bn|$. If $\Om$ is a lattice polygon, construct $\Om'$ by moving each side with Miller normal $\bn_i$ of $\po$ outward by $1/(2|\bn_i|)$, half the interplanar distance. Then extend  the translated sides, so that they once more intersect in the same order as before.  Thus $\Om'$ is a \emph{rational  polygon} \cite{CCD} (not a lattice polygon) that contains the same atoms as $\Om$, with sides  parallel to those of $\Om$ and vertex angles the same as those of $\Om$.  In general though,  it is not a dilation of $\Om$, although $\Om\subset\Om'$. Performing the same operation on $k\Om$ for each $k\in\ZZ$ yields $\Om(k)$. Since the  layers  added to $k\Om$ have measure equal to $k\int_{\po} 1/(2|\bn|) ds$ to dominant order, it follows  from \eqref{remainder} that  the lattice point remainder  $\#(\Om(k)\cap L)-|\Om(k)|=o(k)$. Writing the energy in terms of the modified region $\Om(k)$, one arrives at the following representation:

 \begin{proposition}\label{cure} Let $\ph$ be as in Proposition \ref{infener}, with $d>3$. Suppose $\Om\subset\RR^2$ is (a) a lattice polygon, or (b) a smooth region as in Proposition \ref{smopo}, or (c) the piecewise smooth region of Proposition \ref{conpo}. In case (c) assume further that straight and curved sides of $\po$ are not tangent at their common points. Then for 
 $k\in\ZZ_+$, there exists a convex $\Om(k)\subset\RR^2$ containing the same lattice points as $k\Om$ and whose measure equals the lattice point number of $k\Om$ to order $O(k)$ as $k\to\infty$:
 \be\label{41}\Om(k)\cap L= k\Om\cap L,\qquad
|\Om(k)|=\#(k\Om\cap L) + o(k), \qquad |\po(k)|=|k\po|+O(1),\ee
such that
 \be\label{42}E\{k\Om,y\}=\int_{\Om(k) }W(F) dx +\int_{\po(k)} { \gamma_\circ}(F,n)ds+o(k).\ee
 Moreover, in case (a) the $o(k)$ terms above are $O(1)$, while in cases (b), (c), they are $O({ k^{2/3} })$. Finally, in case (b), \eqref{42} holds with $\Om(k)=k\Om$ and for all $k\in\RR_+$  (not merely integers).
\end{proposition}

\begin{proof} Case (a):
Suppose $\Om$ is a lattice polygon. Then 
$$k\Om=\{x\in\RR^2\colon x\cdot\bn_i\le k d_i,\;i=1,\ldots, N\},$$ is the intersection of $N$ half-planes of the form $ x\cdot\bn_i\le k d_i$, where $\bn_i\in\bbm$ is the Miller normal of the $i$th side and $d_i$ are integers independent of $k$. Let 
\be\label{omko}\Om(k)=\{x\in\RR^2\colon x\cdot\bn_i\le k d_i+1/2,\;i=1,\ldots, N\}.\ee
Thus to construct $\Om(k)$, each straight line containing a side of $\Om$ with Miller normal $\bn_i$ is translated outward (in the direction $\bn_i$) by a $k$-independent distance $1/(2|\bn_i|)$. The intersection of the half-planes of the translated lines is $\Om(k)$ This adds to $k\Om$ a layer whose thickness equals $1/(2|\bn_i|)$ on the $i$th side, hence
\be\label{43}|\Om(k)|=|k\Om|+k \int_{\po}\frac{1}{2|\bn|}ds+O(1)=\#(k\Om\cap L) + O(1).\ee
The $O(1)$ term is a correction due to intersection, in the neighborhood of corners, of layers corresponding to adjacent sides, since directions and thicknesses of layers are $k$-independent. The second equality above follows from \eqref{cpick} of Lemma \ref{pickvar}. The $O(1)$ terms in \eqref{43} are actually constant (depend only on $\Om$ and not on $k$) as is easily shown. This establishes the middle assertion in \eqref{41}. Since the distance of adjacent lattice lines with normal $\bn_i$ is $1/|\bn_i|$, the added layers (whose thickness is half that distance) contain no new lattice points; thus the first assertion of \eqref{41} holds true, while the last is trivial. The first of  \eqref{41}  ensures that $E\{k\Om,y\}=E\{\Om(k),y\}$.
Now \eqref{42} follows immediately from Proposition \ref{infener}, \eqref{43} and the definitions \eqref{se} and \eqref{go}. 

Case (b): Suppose $\Om$ is a smooth region as in Proposition \ref{smopo}. Then choose $\Om(k)=k\Om$, to that \eqref{41} follows from \cite{huxley} and note that \eqref{42} is the same as \eqref{zazen} with $k=r\in\RR_+$.

Case (c): Let $\Om$ comply with Proposition \ref{conpo} . For each $k$ let $\Om(k)$ be the set obtained by moving only the flat sides $k S_i\subset\po_f$, $i\in J_f$ of $k\Om$ outwards by $1/(2|\bn_i|)$ (and discarding portions of the added layers that lie outside the curves $\GG_j$ near the endpoints where $S_i$ join curved sides of $\po$). Thus
\be\label{44}|\Om(k)|=|k\Om|+k \int_{\po_f}\frac{1}{2|\bn|}ds+O(1)=\#(k\Om\cap L) + O( k^{2/3} )\ee
The second equality follows from \eqref{zz1} and \eqref{zz2}. Hence \eqref{41} holds (the last assertion is easy). Once again, \eqref{42} follows easily from \eqref{ffrog} and \eqref{44}.
\end{proof}

\begin{remark}\label{rem8}Proposition \ref{cure} indicates that the  appropriate problem of surface energy minimization over regions of fixed mass  involves minimizing
$ \int_{\po'}\gamma_\circ(F,n) ds$ over a suitable class of domains $\Om'$ with $|\Om'|$ fixed.  The integrand $\gamma_\circ$ is now Lipschitz continuous in the unit normal as guaranteed by Proposition \ref{propg}.  Thus $\gamma_\circ$ can be used to determine the Wulff shape of the crystal.
We must remark, however, that while \eqref{42} has the aforementioned advantages as regards surface energy minimization, it is not appropriate as an asymptotic series in $k$, since the domains of integration depend on the latter variable. The appropriate asymptotic series remains \eqref{muzen}. \end{remark}

\begin{remark}\label{rem7.5}  There is additional motivation for the construction of the  auxiliary domain $\Om(k)$ satisfying the second of \eqref{42}.  Letting each lattice point represent an  atom with unit mass, the total (discrete) mass of $\Om$ is $M(\Om)=\#(\Om\cap L)$. From the continuum viewpoint we expect to be able to write  $M(\Om)=\int_\Om\rho dx$ for some mass density $\rho>0$, which must be  \emph{independent of $\Om$}. Unfortunately, this is not possible for general piecewise $C^1$ domains. Instead we have 
$$M(r\Om)=|r\Om|+O(r)=\int_{r\Om} 1 dx+R(r),\quad  R(r)=O(r)$$
as $r\to\infty$.  The only possible choice would be $\rho=1$ (as expected from the unit atomic mass and unit lattice cell measure), but  the lattice point remainder term $R$ causes problems  since the continuum mass $|r\Om|$ is not equal to the discrete mass unless $R=0$.  This term  is of lower order than the mass, but  of the same order as the surface energy.  In the case of lattice polygons,  $R$ is known explicitly, see \eqref{cpick} and \eqref{remainder}. Based on \eqref{cpick}, one could perhaps  modify the concept of mass density.  One could include a surface mass (second term in \eqref{cpick}) with a corresponding surface mass density, and corner masses (third term). On the other hand for more general $\Om$ it is not even possible to express $R$ explicitly, so the concept of mass density would still be be in question.  It seems the only choice is to somehow choose the continuum domain  properly. In this spirit, Blanc \emph{et al.} assume that there a sequence of dilation factors $r_k$ such that $R(r_k)=0$; this would resolve the mass density issue. Since  the existence of such a sequence is open in general, we construct the auxiliary domains $\Om(k)$ of Proposition \ref{cure}.  One might note that these still involve a lattice point remainder  of order $o(k)$, but this can easily be fixed, for example by adding a suitable term $\e_k$  to the right hand side of the inequality in \eqref{omko}. One then may choose $\e_k$ to eliminate the $o(k)$ remainder so that  $|\Om(k)|=\#(\Om(k)\cap L)$. This eliminates the need for surface and corner masses.  The corrected domains so obtained have a well defined continuum mass density $\rho=1$, while their discrete and continuum masses coincide.\end{remark}

\begin{remark}\label{rem10}  An anonymous reviewer has suggested that it would be advantageous to use the alternative approach of Theil \cite{Theil}, who defines the surface energy as the dominant term of  
\be\label{theil}E\{r\Om,y_r\}-\frac{\#(r\Om\cap L)}{|\Om|}\int_{\Om} W(\nabla y)dx,\ee
 as $r\to\infty$, where $y_r=ry(\frac{\cdot}{r})$ (in case $y$ is not  affine).   We briefly reexamine some of our conclusions from this viewpoint.  If we assume $\Om$ is a lattice polygon and restrict $r$ to integer values, then it follows easily from our results (Proposition \ref{infener} and Lemma \ref{pickvar})  that for affine $y$
\be\label{the2} E\{k\Om,y\}=\frac{\#(k\Om\cap L)}{|\Om|}\int_{\Om} W(F)dx+k\int_{\po} { \gamma_\circ}(F,n)ds+o(k),\ee
$k\in\ZZ$. This is consistent with the alternative approach \eqref{theil}.  The advantage is that the surface energy density in \eqref{the2} is the continuous one $\gamma_\circ$, as opposed to the problematic function $\gamma_\diamond$ in \eqref{infen}. On the other hand, the aim of this paper is to write the discrete energy \eqref{-4} in the canonical form of  continuum mechanics, where the bulk term is $\int_\Om W(\nabla y)dx$ as in  \eqref{-3} or \eqref{-2}, without the  factor  $\#(r\Om\cap L)/|\Om|$.  Moreover, for  general  piecewise $C^1$ domains, the lattice point number $\#(r\Om\cap L)$ cannot be obtained explicitly\footnote{An example of an explicit characterization of $\#(r\Om\cap L)$ in terms of  geometrical aspects of $\Om$  is \eqref{cpick}, valid for lattice polygons.}, hence a drawback of \eqref{theil} is that the bulk energy is not explicitly characterized. The issue is once again with the lattice point remainder  $R(r)=\#(r\Om\cap L)-|\Om|$. When $R(r)$ is not suitably controlled, either the bulk energy in  \eqref{theil}, or the surface energy in our approach, will not be explicitly  determined. Blanc \emph{et al.} write the energy as $E\{r\Om,y_r\}\big/M(r\Om)$, where $M(r\Om)=\#(r\Om\cap L)$ \cite{Blanc}. In order to obtain a surface energy, they also need to control the lattice remainder by assuming a sequence $r_k$ such that $R(r_k)=0$; see \eqref{bbl}. If such a sequence exists, the  factor  $\#(r_k\Om\cap L)/|\Om|$ becomes $1$ and the two approaches coincide. Since it is not known for which $\Om$ this  holds true, our strategy for controlling $R$ is to redefine the continuum domain (keeping the discrete set of lattice points) so that $R    $ is of order lower than the surface energy.  This has the  advantage that the discrete and continuum mass of the body are the same (modulo lower order terms). Once this is done (Proposition \ref{cure}),  the  approach of Theil  and the present one  become equivalent. Specifically, under the hypotheses of Proposition \ref{cure}, for the auxiliary domain $\Om(k)$, the bulk energy in \eqref{theil} reduces to  $\int_{\Om(k)}W(F)dx$ since $\#(k\Om\cap L)=|\Om(k)| + o(k)$; \emph{cf.} \eqref{41}.  The auxiliary domain construction thus eliminates the difficulties associated with the lattice point remainder and yields explicit expressions for both  the bulk and surface energy, in addition to a notion of mass density (see Remark \ref{rem7.5}) in a form consistent with continuum mechanics. 
\end{remark}

\subsection*{Acknowledgments}
 I would like to acknowledge extensive discussions with R.D. James and J.M. Ball, and the hospitality of the Oxford Centre for Nonlinear PDE. I am  grateful for interactions with P. Karageorge, C. Makridakis and D. Mitsoudis. The hospitality and support of the Hausdorff Research Institute of Mathematics of the University of Bonn is also acknowledged.
This research was partially supported by the European Union's Seventh Framework Programme (FP7-REGPOT-2009-1) under grant no. 245749 through the Archimedes Center for Modeling, Analysis and Computation (ACMAC) of the Department of Applied Mathematics at the University of Crete.
Also by  the ARISTEIA  programme  ``Analysis of discrete, kinetic and
continuum models for elastic and viscoelastic response'' of the Greek Secretariat of Research.

\medskip
Received xxxx 20xx; revised xxxx 20xx.
\medskip

\end{document}